# The Anti-Coincidence Detector for the GLAST Large Area Telescope

A. A. Moiseev[1, 2, *], R. C. Hartman[1], J. F. Ormes[1], D. J. Thompson[1], M. J. Amato[1],
T. E. Johnson[1], K. N. Segal[1], and D. A. Sheppard[1]

[1]*NASA's Goddard Space Flight Center, Greenbelt, MD 20771*

[2]*University Space Research Association, Columbia, MD 21044*

*Abstract.* This paper describes the design, fabrication and testing of the Anti-Coincidence Detector (ACD) for the Gamma-ray Large Area Space Telescope (GLAST) Large Area Telescope (LAT). The ACD is LAT's first-level defense against the charged cosmic ray background that outnumbers the gamma rays by 3-5 orders of magnitude. The ACD covers the top and 4 sides of the LAT tracking detector, requiring a total active area of ~8.3 square meters. The ACD detector utilizes plastic scintillator tiles with wave-length shifting fiber readout. In order to suppress self-veto by shower particles at high gamma-ray energies, the ACD is segmented into 89 tiles of different sizes. The overall ACD efficiency for detection of singly charged relativistic particles entering the tracking detector from the top or sides of the LAT exceeds the required 0.9997.

1. Introduction.

The Gamma-ray Large Area Space Telescope (GLAST) is a new gamma-ray observatory scheduled to be launched in 2007. Developed by an international collaboration, including contributions from the U.S. National Aeronautics and Space Administration and Department of Energy, it contains two instruments: the Large Area Telescope (LAT) [1] and the GLAST Burst Monitor [2]. The LAT will detect celestial gamma-rays in the energy range from ~20 MeV to >300 GeV with angular, energy, and time resolution that are substantially better than in the earlier Energetic Gamma Ray Experiment Telescope (EGRET) on the Compton Gamma Ray Observatory [3]. The scientific tasks for LAT originate from results obtained by EGRET and a number of other astrophysical space missions as well as results from TeV ground based gamma-ray instruments. LAT goals cover a wide range of topics: understanding of the mechanisms of charged particle acceleration in active galactic nuclei, pulsars, and supernova remnants, determining the nature of the still-unidentified EGRET sources, detailed study of gamma-ray diffuse emission (both Galactic and extragalactic, as well as that produced in molecular clouds), high energy emission from gamma-ray bursts, transient gamma-ray sources, probing dark matter and the early Universe.

The LAT consists of three main detector systems, a silicon strip tracker, a CsI calorimeter, and an Anti-Coincidence Detector (ACD). The conceptual design of the LAT is shown in Fig. 1. The tracker, in which the gamma rays interact by pair production, provides instrument triggering and determines the arrival direction of detected photons. The calorimeter also provides instrument triggering and measures the energy of detected photons. The ACD surrounds the tracker and provides rejection of charged particles. One challenging feature of the LAT is that it does not have a specially designed directional trigger system, as for example the time-of-flight system in EGRET. The LAT hardware trigger is created by the tracker from the coincidence of signals in 3 consecutive tracker XY layers, or by the calorimeter if the energy deposition there exceeds some pre-selected level. This approach results in very high rate of first level triggers, up to 10 KHz, mainly caused by primary and Earth albedo cosmic rays (protons, helium and other nuclei, electrons), which over the energy range of interest outnumber gamma-rays by 3-5 orders of magnitude. Most of this charged particle background must be removed on-board, prior to transmission of data to the ground, to make the event rate consistent with the available data downlink rate. This requirement makes the task of charged particle identification and rejection one of the main problems in designing the instrument. This responsibility is primarily assigned to the ACD.

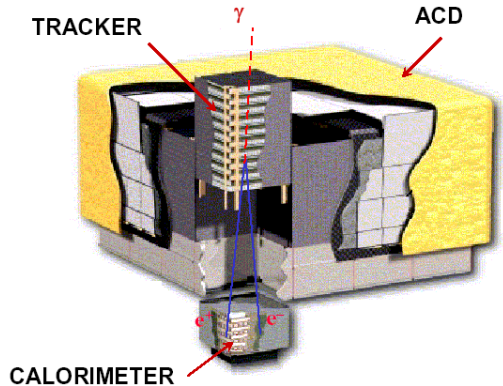

Figure 1. GLAST LAT schematics. The instrument contains 16 identical tracker and calorimeter towers, one of which is shown exploded

2. ACD requirements

2.1 Charged particle detection efficiency

The purpose of the ACD is to provide charged particle background rejection. This purpose dictates its main requirement to have high charged particle detection efficiency. The LAT specification is to have any residual background or "fake photons" at the level of no more than 10% of the diffuse gamma-ray background intensity.

Fig. 2 compares the differential spectra of cosmic ray protons [4] and electrons [5] to that of the extragalactic diffuse gamma-ray background [6] (extrapolated beyond the measured limit of 120 GeV). These spectra are shown with an approximation of the geomagnetic cutoff for the GLAST low-Earth orbit with 28º inclination and altitude of 565 km (orbital decay has little effect). Also shown is the required background level, 10% of this diffuse flux. In order to meet this requirement, LAT needs a factor of ~$10^6$ suppression of protons and ~$10^4$ suppression of electrons. Proton rejection is provided by the ACD along with the tracker and the calorimeter. Calculations indicated that the LAT calorimeter and tracker provide proton suppression by at least a factor of $10^3$, employing event patterns in the tracker and shower shapes in the calorimeter. The ACD must provide the remaining factor of $10^3$.

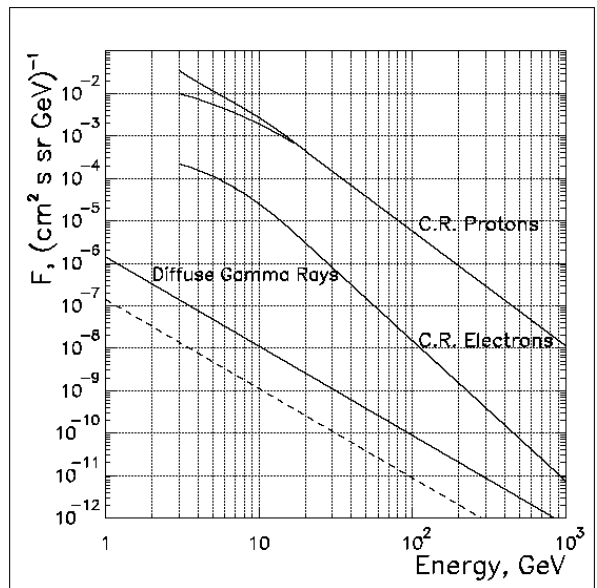

Figure 2. Differential spectra of primary cosmic rays. Solar modulation effect is shown for energy below 15 GeV; for protons the maximum and minimum solar modulation are shown. The dashed line shows 0.1 of the extragalactic diffuse gamma-ray flux. The extragalactic diffuse gamma-ray spectrum is extrapolated beyond the upper range of 120 GeV measured by EGRET

Since the cascades of electrons and gamma rays will appear very similar in the calorimeter, the ACD and the tracker will be the primary tools against electrons. The electron spectrum is steeper than the photon spectrum, so for lower energies the requirement is tighter. At 3 GeV, approximately the lowest energy for which primary cosmic ray electrons can reach the LAT in its orbit, the requirement becomes that no more than 1 in ~$3 \times 10^4$ electrons can be classified as a gamma ray in the LAT. This is equivalent to an electron detection (recognition) efficiency of 0.99997 for the LAT.

The tracker can help in rejecting this electron background, but this rejection comes at some expense of photon detection efficiency. The tracker can be used as an anti-coincidence barrier by requiring that the pair production conversion layer be clearly identifiable by the absence of a signal in the outer layer or layers, projecting backwards the path defined following the conversion. Events that project directly back to inefficient zones in the ACD can be identified and "cut out" of the data sets. We estimate that this approach can provide charged particle suppression by at least 10. Nevertheless, electrons represent the most challenging background to remove, and drive the efficiency requirements for the ACD. As a result of these considerations, the ACD is required to provide at least 0.9997 efficiency (average over the ACD area) for detection of singly-charged particles entering the tracking detector from the top or sides of the LAT.

## 2.2. Suppression of the backsplash effect

The LAT is designed to measure gamma-rays with the energies up to at least 300 GeV, where the detector will begin to run out of photons due to the rapidly falling spectra of cosmic gamma-ray sources. The requirement to measure photon energies at this limit leads to the presence of a heavy calorimeter (~1800 kg) to absorb enough of the energy to make this measurement. The mass itself, however, creates a problem we call the *backsplash effect*. A small fraction of secondary particles (mostly 100-1000 keV photons) from the electromagnetic shower, created by the incident high energy photon in the calorimeter, travel backward through the tracker and cross the ACD, where they can Compton scatter and thereby create signals from the recoil electrons. These ACD signals will be interpreted by the instrument as vetoes, and otherwise legitimate high energy incident photon events could be rejected (Fig. 3). This effect was present in EGRET, where the instrument detection efficiency for 10 GeV photons was a factor of two lower than at 1 GeV due to false vetoes caused by backsplash. At energies above ~50 GeV EGRET was almost insensitive due to this effect [7]. For the LAT we established a design requirement that vetoes created by backsplash (self-veto) would reject not more than 20% of otherwise accepted photons at 300 GeV. But the area of the LAT ACD is large enough that a monolithic (unsegmented) single detector would render LAT "blind" to photons of such a high energy.

In order to understand how to prevent serious problems from this effect, we carefully studied the probability of backsplash [8]. That paper gives an empirical formula for backsplash "hit" probability per unit area, as a function of energy and distance backwards from the shower. We determined that at the distance from the cascade where the ACD would be placed, at 300 GeV there was a ~10% probability per 1000 cm$^2$ of a backsplash Compton photon signal in the ACD.

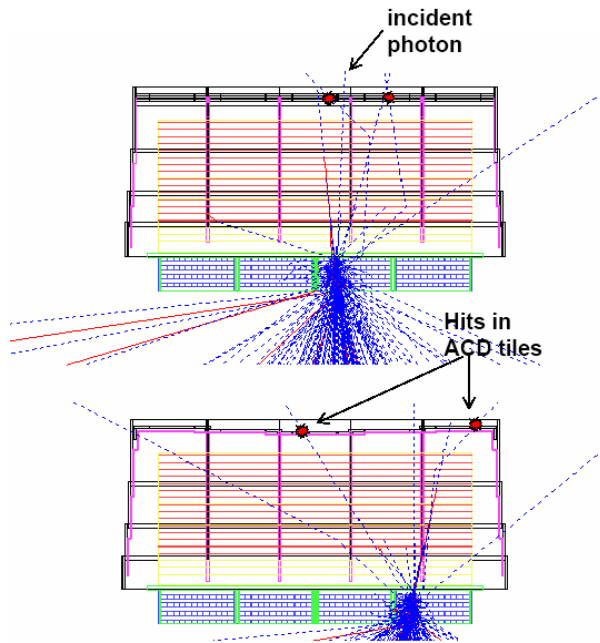

Figure 3. Backsplash in the LAT ACD simulation model. Charged particles are shown by red lines, and protons by blue dashed lines. Signals in the ACD caused by backsplash are shown by red dots

We established that *the way to suppress the backsplash effect is to segment the ACD,* and to consider in analysis only the ACD segment in the projected path of the incident photon. This approach dramatically reduces the area that can contribute to backsplash, from the 83,000 cm$^2$ total area of the ACD to the area of the largest segment, ~1000 cm$^2$.

Another problem is that, in order to achieve the required high detection efficiency for singly-charged minimum ionizing particles (hereafter *mips*), the ACD signal detection threshold must be set as low as possible because of statistical fluctuations of the energy loss in the scintillating detector (assumed to follow the Landau distribution). On the other hand, backsplash signals are mainly Compton electrons from soft photons, and they create a characteristic soft signal spectrum in the same detector (which at 1 cm is thick enough to absorb most of these low energy Compton electrons). So in order to reduce the backsplash-caused self-veto signals, the detection threshold needs to be set high. These two requirements on signal detection are in competition, as illustrated in Fig. 4. This means that the ACD design must be carefully optimized, and the event veto threshold setting must be carefully tuned. For example, the veto threshold must be constant over the entire ACD area, which requires the detected light signal also to be constant, and the natural fluctuations to be minimized, driving the requirement for scintillation photons to be collected and turned into photoelectrons efficiently. These requirements set strict limitations on the choice of particle detectors and readouts to be used for the ACD, as will be discussed in section 3.

2.3. Total amount of ACD material

Two issues constrain the material thickness of the ACD:

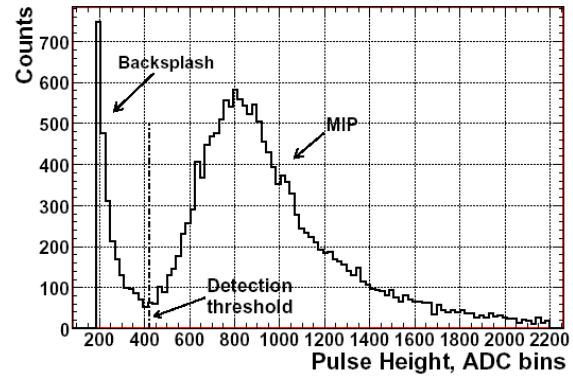

Figure 4. Conflict between efficient charged particles detection and backsplash suppression. Pulse-height distribution of signals from an ACD tile is shown, with mip peak and backsplash spectrum

a) The ACD is an absorber for the incoming gamma radiation to be detected by the LAT. In planning the design, we required that the ACD, micrometeoroid shield (MMS) and thermal blanket together (along with their associated mounting hardware) must not cause interactions (either pair production or Compton scattering) of more than 6% of the incident gamma rays at energies in the LAT range.

b) The ACD provides target material for cosmic rays to produce gamma rays in the LAT energy range, largely through neutral $\pi^0$ production/decay and positron annihilation. If such gamma-ray-producing interactions do not also produce a signal in the ACD itself, they generate a local background essentially indistinguishable from cosmic gamma rays. In particular, the inert material outside the active ACD detectors must be minimized, consistent with protection required against the space environment. The goal was to make the thermal blanket/micrometeoroid shield no more than twice the thickness of the EGRET system, i.e. 0.34 g/cm$^2$. The larger thickness is required by the larger area of the LAT ACD and the increase in orbital debris since the CGRO mission (for details, see section 3.9.2).

2.4. Design constraints

As a subsystem of a space flight instrument, the ACD had to be built within allocated limits. These limitations include the following:

a) The ACD must comply with the vibration, acoustic, and acceleration requirements for an instrument to be launched on a Delta-II vehicle, and with the normal restrictions on materials used in space-flight instruments;

b) The ACD must have 95% probability of performing within specifications for 5 years of operation on orbit, with the exception of the loss of one ACD tile due to micrometeoroid penetration (see section 3.9.2).

c) The mass of the ACD must be less than 290 kg and its orbital average power less than 12 W.

d) The ACD must fit between the LAT tracker and the launch vehicle shroud. The outer dimension of the ACD and consequently the LAT is limited by the launch vehicle shroud. Because the LAT tracker must fit within the ACD, minimizing the thickness of the ACD allows us to increase the footprint of the LAT tracker, one of the key LAT parameters.

e) The operational temperature range of the ACD is: −40 C to +40 C (detectors); −20 C to +40 C (electronics). The relatively wide range of operational temperature implies a requirement to have sufficient gaps between ACD elements to tolerate thermal expansion and contraction, as well as the vibrations occurring during the Delta-II launch. It will be shown in section 3.12 that these gaps have a significant impact on ACD design.

f) The ACD must cover the top and all four sides of the tracker, requiring a total ACD area of ~1700 mm × 1700 mm on the top, and ~1700 mm × 800 mm on each side.

3. ACD design - issues and implementation.

In this section we describe the top-level ACD design drivers and solutions, list trade-off issues considered, and discuss in more detail the most important design issues.

3.1. Design overview

Three main requirements must be considered together: high efficiency for charged particle detection, adequate tolerance to backsplash, and reliability. Numerous trade studies and tests have been performed in order to optimize the design, in laboratory tests, two beam tests [9], a balloon flight [10], and comprehensive simulations. As a result, the basics of the design are as follows (see Fig. 5):

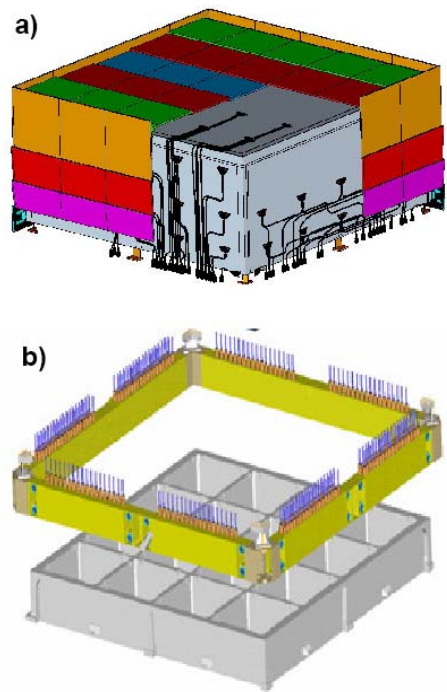

Figure 5. ACD Structure. a) – ACD tile shell assembly, with tile rows shown in different colors. Clear fiber cables are seen in the cutout. Ribbons and bottom row (long) tiles are not shown. b) – ACD base electronics assembly (yellow) with PMTs shown. The LAT grid is shown on b) in gray

1) The ACD is made of *plastic scintillator* detectors, chosen as the simplest, most reliable, efficient, well-understood, inexpensive, and robust detector technology, with much experience in space applications;

2) The ACD is *divided into segments (hereafter called "tiles") in order to suppress the backsplash effect*. Although there are some operational complexities (see sections 3.2 and 4), the basic principle is that an event is vetoed only if the reconstructed particle trajectories from the tracker extrapolate to an ACD tile with a measurable signal. This combination of tracker and ACD information reduces the self-veto rate by nearly two orders of magnitude. Also, with every ACD tile separately light tight, a puncture by a micrometeoroid can disable only one tile, causing system performance degradation by no more than a few percent, which is tolerable;

3) *Wavelength shifting fiber (WLS) readout* of the scintillator light provides uniformity of light collection over each detector segment, and allows photomultiplier tubes (PMTs) to be placed well away from the scintillator tiles, reducing the amount of material in the instrument aperture and simplifying instrument assembly and repair (if needed);

4) Overall detection efficiency for incident charged particles is maintained by *overlapping scintillator tiles in one dimension*. In the other dimension, gaps between tiles are covered by *flexible scintillating fiber ribbons (section 3.12)*. The ribbons follow the gaps between tiles and provide detection of particles that enter through the gaps;

5) The ACD reliability is increased by using *two redundant PMTs to sense the light from each scintillator tile (section 3.14)*;

6) All ACD electronics and PMTs are positioned around the bottom perimeter of the ACD, and light is delivered from the tiles and WLS fibers by a *combination of wavelength-shifting and clear fiber cables (Section 3.13)*;

7) The ACD electronics is divided into 12 groups of 18 channels, with each group on a single circuit board (section 3.11). Each of the 12 circuit boards, known as FREE (FRont End Electronics) boards, is independent of the other 11, and has a separate interface to the LAT central electronics. Each FREE board services signals from 15-17 PMTs, located adjacent to the board. The PMTs associated with a single FREE board are powered by a High Voltage Bias Supply (HVBS), with redundant HVBS supplies for each board. For reliability, the two light signals from a tile are always routed to PMTs associated with different FREE boards;

8) In order to reduce the gaps between tiles and minimize the amount of inert material, *each ACD tile is mounted without a frame, and is wrapped in two layers of high reflectance white Tetratec, then by two layers of light-tight black Tedlar*;

9) To minimize the chance of fatal light leaks due to penetrations of the light-tight wrapping by micrometeoroids and space debris, the ACD is completely surrounded by a *micrometeoroid shield* (MMS), described in section 3.9.2;

10) At grazing incidence, the large top flat portion of the MMS provides long path lengths through inert material for hadronic cosmic rays to produce neutral pions ($\pi^0$), which immediately decay into two gamma rays. Some of those gamma rays can pass through the ACD undetected and form a background indistinguishable from cosmic gamma rays. This background from cosmic-ray $\pi^0$ production is minimized by *extending the top row of side tiles above the tiles in the ACD top to the upper surface of the micrometeoroid shield* (MMS). This forces charged products of a grazing $\pi^0$-production event to pass through and be detected in a scintillator tile. This extension is known as the "crown".

The design outlined above and described in detail in later sections of this paper is the result of many tests and trade studies. Although the details of all these studies would be beyond the scope of this paper, we list here some of the alternatives that were considered:

- Single or redundant tiles for each segment (two layers);

- ACD segmentation – the number of segments and how to deal with the necessary gaps between tiles (e.g., tiles might be glued together but optically separated);

- Suitability and geometry of ribbons to cover gaps;

- Overlapping tiles in one dimension or two;

- Light delivery from tile to PMT – whether to run WLS fibers the whole length, or use clear fiber extensions optically mated to the WLS fibers to reduce light attenuation;

- Whether to position the PMTs near the tiles or to have them grouped at the base. This decision drove the high voltage and readout electronics organization;

- Functions and requirements for the ACD electronics: e.g., does it need to include part of the LAT trigger logic, or should that function be carried out by the central LAT electronics;

- Numerous optimization studies in the tile design, including tile thickness, WLS fiber attachment, and groove pattern.

### 3.2. ACD Veto organization

The ACD provides charged particle rejection in three ways:

a) ACD signals are used on-board as a "throttle" on the first level (hardware) trigger. Combinations of ACD tiles are defined to "shadow" each of the 16 tracker towers. The ACD fast discriminator veto signals from these tiles are included in the first-level trigger to reduce the number of triggers that must be processed by the on-board software.

b) For on-orbit event filtering, the main task is to reduce the event rate due to charged cosmic rays to a level compatible with the data transmission rate to the ground. (For calibration purposes, it is useful to transmit a very small fraction of the charged particle events to the ground.) The expected maximum LAT first level trigger rate is ~10 kHz, and the average downlink rate (which depends on the event size) is around 300 Hz, of that 4-5 Hz are the gammas. On-board processing compares the tracker and calorimeter information with the ACD signals in order to discard those events that have a high probability of being charged particles. At this point ACD must provide ≥0.99 efficiency for charged particle detection. Signals for this purpose are obtained from onboard discriminators, with veto thresholds that can be higher than would be needed to achieve the ultimate goal of 0.9997 efficiency.

c) The ultimate efficiency is obtained during off-line analysis on the ground. For this purpose the discriminator and pulse height analysis (PHA) signals from every ACD PMT are digitized on-board and telemetered with every LAT event sent to the ground. The ACD data represent a small fraction of the total event size and have no significant impact on the telemetry rate. The needed thresholds are determined from off-line analysis of PHA distributions of signals created by normally incident *mip*s for every PMT.

### 3.3 Scintillating tile detectors

The Tile Detector Assemblies (hereafter TDAs) are the central detecting elements of the ACD; a detailed description is given in [11] along with performance details. They are made of ElJen-200 plastic scintillator, 10 mm thick (everywhere but in the central row of TDAs on the top of LAT, where the thickness is 12 mm; see section 3.4.e for details). Scintillating light from all tiles is collected by 1 mm diameter wavelength shifting (WLS) fibers BCF-91A, made by Saint-Gobain (formerly Bicron). The ElJen-200 scintillation light emission peak is at 425 nm, which matches the BCF-91A maximum absorption wavelength, providing maximum

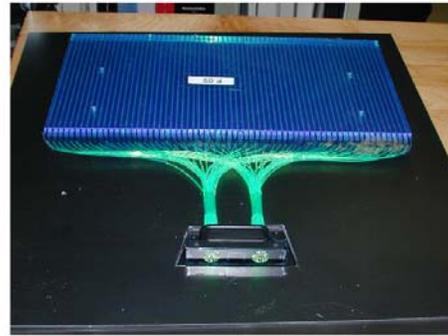

Figure 6. Scintillator tile detector assembly (TDA), shown unwrapped. The green wavelength shifting (WLS) fibers carry light to the optical connector in the foreground

light conversion for transmission to the PMTs. The choice of materials for the TDAs was driven largely by matching the scintillator emission to the WLS fibers absorption. Because the scintillator has no special requirements for timing or long absorption length, we chose the ElJen-200 as a reliable, low-cost, rugged material. The dimensions of the TDAs vary from 15 cm by 32 cm (side 3rd row) to 32 cm by 32 cm (top of ACD). The total number of these tiles is 85. All have similar design (Fig. 6). In addition, there is a long tile, 17 cm by 170 cm, at the bottom of each side (see sections 3.6 and 3.8 for details of this decision).

The TDA design provides a highly efficient and uniform collection of scintillation light over each tile area by using the WLS fibers. High light yield is needed to achieve the required charged particle detection efficiency, because the detection inefficiency is determined by signal fluctuations. The main contribution to this fluctuation comes from the smallest number in the chain of detection processes, the number of PMT photoelectrons created by the light from the scintillator and WLS fiber. It was determined that a detection efficiency of 0.9997 at a detection threshold of 0.3 *mip* implies that a *mip* must produce about 20 photoelectrons in a PMT (Fig. 7). This estimate gives a lower bound for the light yield needed to meet the overall ACD efficiency requirement. To increase the detector reliability, each tile is read out by two PMTs (and the *mip* must generate ~20 photoelectrons in *each* PMT.)

3.4. Maximizing production, collection, and transmission of light

Since the effective light yield (hereafter light yield, but incorporating scintillator light yield, collection efficiency, transmission effects, and PMT conversion efficiency) from the scintillator tiles is crucial in achieving the ACD efficiency requirement, significant efforts were made to increase it [11]:

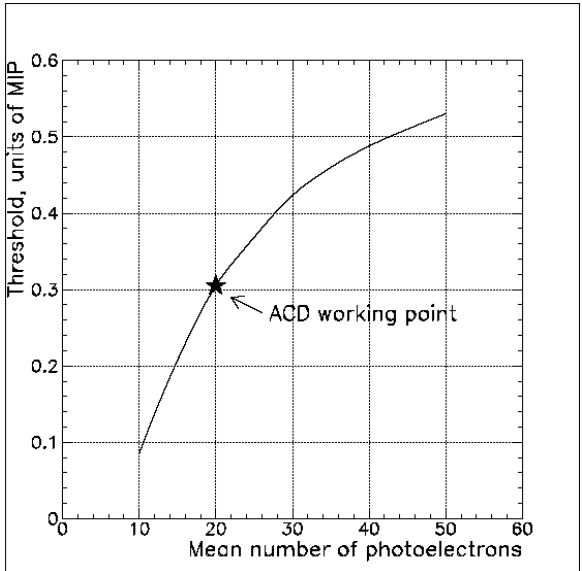

Figure 7. Mean number of photoelectrons needed to provide 0.9997 detection efficiency by a single tile vs. the discriminator threshold

a. It was found that WLS fiber BCF-91A is the best match to the scintillator used;

b. The depth of fibers embedding in the tiles is 2 mm; their pitch is 5 mm, with fibers alternately routed to two PMTs;

c. Fiber ends are polished, and those ends inside tiles are also aluminized, via vacuum deposition;

d. Scintillators are wrapped in two layers of Tetratec (250 μm thick), found to be the best light reflector for our detector;

e. Scintillator tile thickness was chosen as a compromise between performance and mechanical constraints (mass and volume), and is 10 mm everywhere except for the tiles in the middle row of the ACD top surface. These five tiles have the longest light paths to their PMTs, and consequently the largest light loss in transmission (discussed later), so their thickness is 12 mm.

Light yield measurements have been performed many times throughout the TDA design process. Two approaches were used to measure the number of photoelectrons created at the PMT photocathode. In the first, the PMT was calibrated with a light-emitting diode (LED) to get a signal from a single photoelectron; after that the signal obtained from normally incidence *mips* was expressed in units of photoelectrons. A second approach, most frequently used in our work, was based on measuring the scintillating tile efficiency for detecting *mips* (cosmic-ray muons) at different detection thresholds, then fitting the result with a Poisson distribution for different numbers of photoelectrons. The principal difficulty here is to select a very clean sample of muons to be used for the tile efficiency determination. The experimental setup used to detect muons by an ACD tile is shown in Fig. 8, where the tile being tested is placed between two triggering scintillator tiles of smaller size, S1 and S2. Two 5 mm thick lead absorbers F1 and F2

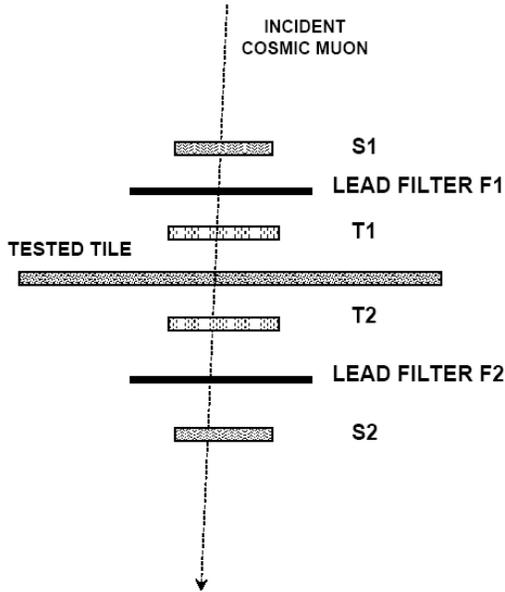

Figure 8. Experimental setup for tile efficiency determination using cosmic muons

are used to remove the soft component cosmic rays. Signals from scintillators T1 and T2 are used to select *mip*s by analyzing their pulse heights. An example of the result is shown in Figure 9. In this approach we obtain an effective number of photoelectrons, which actually is an underestimate of the true number, because we neglect the contribution from particle ionization loss fluctuation in the scintillator (Landau fluctuation) and electronics contributions to the width of the observed signal distribution. However, the estimate obtained for the photoelectron number reflects actual detector efficiency, and we take it as a figure of merit for the scintillator tile performance. The best Poisson fit to the experimental curve in Fig. 9 corresponds to $N_{p.e.}=23$, which we take as the true value for this particular detector/PMT combination. It is important to note that the light yield of the detector depends on five major contributions for a given energy deposition in the scintillator: 1) the amount of light produced in the scintillator (depends on the scintillator material); 2) efficiency of light collection from the scintillator to the WLS fibers; 3) efficiency for conversion of scintillator light to wavelength-shifted light (depends on both the scintillation light spectrum and the WLS material); 4) efficiency of light transmission to the PMT; and 5) quantum efficiency of the PMT photocathode. We discussed in section 3.3 the choice of the scintillator and light collection efficiency. Light transmission from the tile to the PMT is discussed in section 3.13. PMT quantum efficiency, which is a property of the PMT photocathode, is the efficiency of light photon

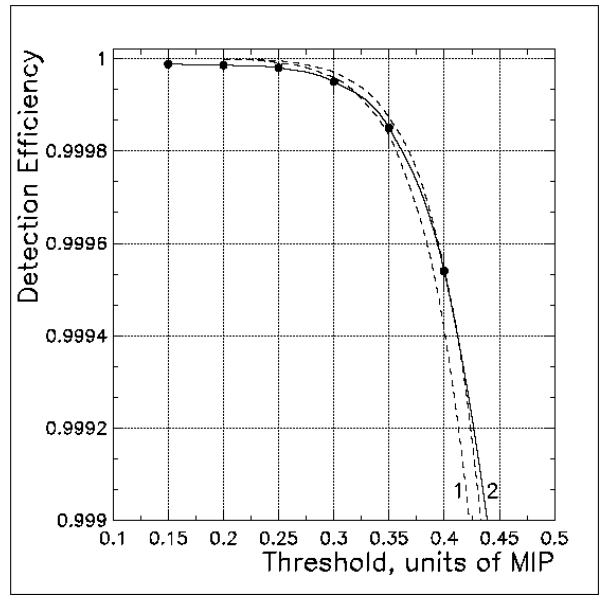

Figure 9. Example of tile light yield determination. Solid line shows measured efficiency, and dashed lines correspond to the Poisson distribution with 22 photoelectrons (line 1) and 24 photoelectrons (line 2)

conversion into photoelectrons (normally around 15-20% for the light generated by the WLS fibers, ~490 nm wavelength). We take into account the variations in PMT quantum efficiency in all results presented here.

### 3.5. Light collection uniformity over a tile

Because of the conflict in the choice of optimal detection threshold between achieving maximum detection efficiency and suppression of the backsplash effect, the uniformity of light yield over the tile area is very important. It needs to be within 10% of its average value, excluding the tile edge area, in order to allow a uniform detection threshold over the tile area, and consequently uniformity of particle detection efficiency and backsplash suppression. The use of WLS fibers appears to provide the best uniformity in light collection, and fiber spacing was optimized to achieve that uniformity. In order to map tile light yield uniformity we used two plastic scintillating hodoscopes specially designed and built for these tests. Each hodoscope detector contains 8 plastic scintillator strips, each 4 cm wide, and the two hodoscopes are oriented with their strips orthogonal to each other (X and Y), with one hodoscope above and the other below the tile being tested. Atmospheric muons provide the source of minimum ionizing particles. The signal from the tile being tested is analyzed when coincident signals are received from both upper and lower hodoscopes. In analysis, by using the coincidences between a single strip in each hodoscope, positions are mapped out on tile being tested. Thus a light yield map can be created with 4 cm × 4 cm pixels. The goal is to achieve light yield uniformity such that the signal average from any tile pixel is within 10% of the signal averaged over the whole tile. In this test, it was found that the light collection is acceptably uniform in the middle of the tile, but that there is 20-30% light yield reduction in the edge pixels, caused by light escape at the scintillator edges. This edge effect was further explored using a more precise 1 mm scintillating fiber hodoscope for event selection. This showed that the light yield decreases to approximately 70% at the tile edge, and recovers to 100% at ~3 cm from the tile edge. Simulations show that this effect reduces the overall ACD efficiency by $\sim 5\times 10^{-5}$.

More precise and reliable measurements were performed in muon tests of the full LAT, where the tracker was used to determine the point of muon passage through a tile, with an accuracy of ~0.2 mm. The results are given below in section 5.2.2. The light collection uniformity achieved allows the signal detection threshold, and consequently the particle detection efficiency, to meet the required level of uniformity over the tile area.

### 3.6. Long (bottom row) tile design

The bottom 15 cm of the ACD is not segmented. Like the other ACD tiles, these bottom tiles are used to reject charged particle background. Unlike the other tiles, however, these tiles are outside the LAT primary field of view. No events entering through these tiles will be accepted as gamma rays; therefore the small segmentation needed to avoid backsplash is not required. On each of the four sides, a single-piece of scintillator is used, ~170 cm long and 17 cm wide (to overlap with the row above), with embedded WLS fibers along the whole tile. The readout is provided from near both tile ends, with the fiber grooves bent inside the tile, but exiting through the large tile surface rather than the ends. This is necessary because of the proximity of the corresponding tiles on the two adjacent sides. The light yield from this tile design is lower, and also not as uniform, as in the smaller tiles, due to significant light attenuation in WLS fibers. However, the efficiency requirement for this tile is relaxed to 0.9990 (averaged over the whole tile), which makes the design feasible. Prototype performance tests demonstrated that it meets the required efficiency easily, even in the case of failure of one PMT.

### 3.7. The "crown" tiles

As was mentioned in section 3.1, the top row tiles on all four sides are extended to the top of the MMS and thermal blanket, creating the "crown", in order to suppress potential background created in the MMS and thermal blanket. In this design the MMS and the thermal blanket are shielded from the sides by the extensions so that the protons that could potentially cause background gammas must cross the tile extensions and thus can be vetoed. A simulation study demonstrated that without the crown such background gammas could be as much as 5% of the extragalactic diffuse radiation at energies above several GeV due to this effect. In principle, similar extensions could be useful on the sides. Not only was there essentially no space in which to put such a side crown, but finding a tile shape geometry that would work would have further increased an already complex fabrication.

### 3.8. Segmentation optimization

The key driver for tile size was the backsplash effect. Using the approach described in [8] and the requirement of <20% efficiency reduction due to backsplash, with 0.3 *mip* threshold, for 300 GeV incident photons, the initial estimate for tile size on the top of the ACD was ~1,000 cm$^2$. Using that size, it was decided to divide the top of the ACD into a 5 × 5 array. This puts the interior corners of the tiles roughly above the centers of the 4 × 4 array of tracker towers below. The size of the side tiles decreases toward the bottom of ACD in order to maintain the solid angle of the tiles as seen from the shower maximum in the calorimeter. We used the formula developed in [8] along with full Monte Carlo simulations in the process of segmentation optimization:

$$P_{backsplash} = 2.5 \times \exp\left(-\frac{T_{thr}}{0.19}\right) \times \frac{A}{432} \times \left(\frac{55}{x+15}\right)^2 \times \sqrt{E}$$

where $P_{backsplash}$ is the fraction of photon events (with energy $E$ in GeV) accompanied by signal above threshold $T_{thr}$ (in *mips*) from a tile with area $A$ ( in cm$^2$) at a distance $x$ (in cm) from the calorimeter face.

The ACD consists of 89 tiles, with every tile read out by 2 PMTs, giving 178 electronic channels not counting those used by ribbons. The original segmentation design assumed 145 tiles, with finer segmentation in the two bottom rows of tiles, in order to make use of the events entering LAT through the bottom part of ACD at large angles, and having long paths in the calorimeter (which would have better energy resolution at high energy). But due to strict resource limitations this design was simplified to the 89-tile design slightly reducing the capability to detect large angle events. The effect on the backsplash suppression of having only 89 tiles was investigated and it was found that, for the events entering LAT above the bottom ACD tile row (long tiles), the effective area and geometrical factor decreases by less than 5% at 300 GeV.

Fig. 10 shows the estimated backsplash for 300 GeV incident gamma-rays, entering the LAT through the top, based on the studies described in [8]. It is approximately the same for events entering through the side, because the 2-nd and 3-rd side row tiles (where the distance to the calorimeter can be shorter) are smaller. The expected effect is lower (better) than required, but this analysis was performed assuming the use of only one tile crossed by the reconstructed track, and perfect tracker pointing to the tile. Use of ACD in on-board event selection may include more than one tile in event vetoing (near tile edges and corners), so it is reasonable to have margins.

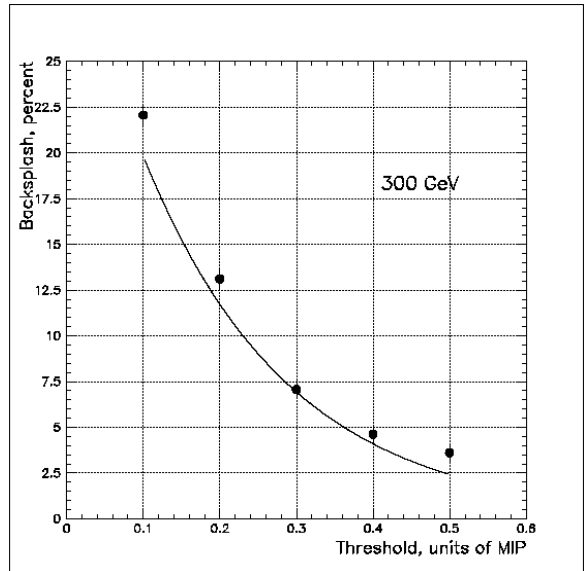

Figure 10. Expected backsplash for LAT for 300 GeV gamma rays entering the instrument through the top [8]. Data points are as measured in beam test, and the line shows the prediction by the formula

### 3.9. Mechanical design

The ACD Mechanical Subsystem consists of two primary assemblies, the Base Electronics Assembly (BEA) and the Tile Shell Assembly (TSA); see Fig. 5. The BEA is the mechanical support structure for the ACD. It supports the TSA, houses the ACD electronics, and provides mechanical and electrical interfaces to the LAT. It is an aluminum structure and is mechanically symmetric (square). The TSA supports the Tile Detector Assemblies (TDAs) and their associated fiber cables as well as the Micrometeoroid Shield/Thermal Blanket. The 24 electrical cables from the LAT are mated to the BEA. The total mass of the ACD is 284 kg. The calculated fraction of gamma radiation absorbed by the ACD is 4.9% on the top and 4.6% on the sides, well under the required maximum of 6%.

The primary advantages of the ACD mechanical design are that it provides a clean and simple interface to the LAT, and it allows the TSA and BEA to be integrated and tested in parallel.

### 3.9.1. Tile shell assembly

The TSA consists of the following components: a composite shell, 8 titanium flexures that attach the TSA to the BEA, 89 TDAs and their associated fiber cables and cable tie-downs, 8 fiber ribbons, and TDA tie-downs.

There are 25 TDAs on the top and 16 TDAs on each of the four sides. On the top, there are two different types of TDAs, 15 flat and 10 bent. The bent TDAs are positioned along two edges of the top; they are required to provide routing access for the WLS fibers from the edge tiles. The 16 side TDAs are arranged in four rows with five TDAs in each of the upper three rows and a single long TDA for the lowest row. The TDAs are "shingled" in one dimension to eliminate gaps in that dimension; the gap in the other dimension is covered using scintillating fiber ribbons, with PMTs on each end (see section 3.12 and Fig. 13).

An important issue was how to mount the tiles to the ACD structure. The frameless mounting requirement drove the decision to attach each tile via 4 screws, which go through 3.2 mm holes in the scintillator. Soft black viton gaskets provide cushioning and light-tightness. The impact of the mounting holes on tile efficiency is addressed in section 3.15.

There are a total of eight ribbons, four to cover the gaps along the X-axis and the other four to cover the gaps along the Y-axis. The entire system is covered by a thermal blanket and micrometeoroid shield.

The composite shell is constructed using honeycomb panels bonded together at the edges. The honeycomb panel has aluminum core with 0.5 mm thick M46J/RS-3 carbon fiber composite facesheets. The top panel is 51.8 mm thick and the side panels are 26.4 mm thick. The 8 titanium flexures are all single blade flexures, oriented with their long dimensions normal to the direction to the ACD center of mass. The flexures are required because of the coefficient of thermal expansion difference between the composite shell and the base frame. The TDA tie-downs provide the interface between the composite shell and the TDAs. They are bonded to the composite shell and provide compliance for the thermal expansion and contraction of the TDAs.

### 3.9.2. Micrometeoroid shield and thermal blanket

As the outermost element of the LAT, the ACD requires special protection from the space environment. In particular, the light-tight wrapping of the TDAs must be shielded from micrometeoroids and space debris, whose high velocities (typically 10 km/s) make even small particles highly penetrating. At the same time, the passive material outside the scintillators must be minimized, because cosmic ray particle interactions with this material can produce gamma rays that become an irreducible local background for the LAT. In consultation with the Hypervelocity Impact Technology Facility team at Johnson Space Center, we developed a Micrometeoroid Shield (MMS) based on the multi-shock shield concept [12]. The MMS, with a total area density of 0.39 g/cm$^2$ (about 10% larger than our design goal), uses four layers of Nextel$^{TM}$ ceramic fabric separated by 4 layers of 6 mm thick Solimide$^{TM}$ low-density foam and backed by 6-8 layers of Kevlar$^{TM}$ fabric. Calculations and tests with a light-gas gun show that the MMS can stop aluminum particles with diameter up to 2 mm traveling at 7 km/s. Based on the NASA ORDEM2000 orbital debris model [13], the MMS has a 95% probability of allowing no more than 1 penetration in five years. Although a single penetration would cause the overall ACD efficiency to fall below 0.9997, simulations show that a single isolated "dead spot" would produce only a few percent loss of overall LAT performance, because the analysis of tracker data could eliminate suspect events entering through this region. For thermal control, the entire ACD is wrapped in a multi-layer insulation (MLI) thermal blanket with an outer layer of germanium-coated Kapton$^{TM}$ to reduce degradation by atomic oxygen in orbit.

### 3.9.3. Mechanical analysis

A NASTRAN$^{TM}$ Finite Element Method (FEM) mathematical structural model was developed for the ACD. The FEM, which contains 8,528 elements and 8,899 nodes, was used to determine stress margins and displacements under inertial loads and to obtain the natural modal frequencies of the system. The TSA, the BEA that supports the TSA, and the flexures were modeled and showed positive margins of safety. The dynamic analysis demonstrated that the ACD system met the requirement of 50 Hz minimum resonance frequency. The first frequency of the GLAST ACD is 54.17 Hz, which corresponds to the lateral mode of the TSA. The displacements of the system, in response to the static loads, were also within the requirements.

## 3.10. Thermal design and analysis

The ACD electronics, the PMTs, and the scintillator tiles are temperature sensitive. The ACD contains twelve electronics boards that are mounted to the BEA frame, each with power dissipation of less than 1 W, for a total ACD power dissipation of 11.4 W. The ACD PMT assemblies dissipate no significant power. The electronics operating temperature range is –20 C to +35 C, while the survival range is –40 C to +45 C. The scintillator tiles have no power dissipation. Their operating temperature range is from –30 C to +35 C. The survival range is –60 C to +45 C. The thermal design strategy to satisfy the ACD electronics board temperature requirements was to utilize the LAT aluminum grid structure (operating in a temperature range –10 C to +25 C) as a heat sink for the electronics boards. The temperatures of the ACD TDAs are attained by balancing the heat flow radiated from the tracker with the heat flow through the MMS and MLI blankets. The effective ACD thermal emittance to space is approximately 0.01.

Two orbit cases are used for analysis and temperature predictions, taking into account the solar radiation, the earth infrared and the solar albedo inputs and their effects on the ACD and the LAT grid (which is cooled by radiators). The first case is the hot case orbit, where either of the ±Y faces (the sides with the radiators) is in the sun for the duration of the orbit, and the second case is the cold case, where either of the ±X faces is in the sun (both radiators facing space). In both cases the +Z axis is anti-nadir pointing. For the "point anywhere, anytime" operation of this mission, these two cases bound the possible ACD temperatures.

## 3.11. Electronics design

All ACD electronics are situated in eight electronics chassis mounted in the Base Electronics Assembly, which consists of:

- 194 PMTs (two per tile plus two per ribbon)
- 12 front-end electronics (FREE) boards, each handling 15-17 phototubes.

- Two (redundant) High Voltage Bias Supplies (HVBS) with each FREE board.
- 24 electrical interface connectors

The GLAST LAT ACD electronics system is used to provide charge collection, amplification, and signal processing for the ACD scintillating tiles. The ACD interfaces to the LAT GASU (Global Trigger/ACD Electronics Module (AEM)/Signal Distribution Unit). The output of each photomultiplier tube is used for two purposes: 1) generation of veto pulses, which are used by the LAT trigger electronics to eliminate nearly all of the events triggered by charged particles; and 2) after shaping (4 μs), measurement of the signal amplitude for transmission to the ground, where precise analysis is carried out for the final charged particle rejection. The data are formatted and transmitted to the LAT computer via the GASU for event filtering and downlink.

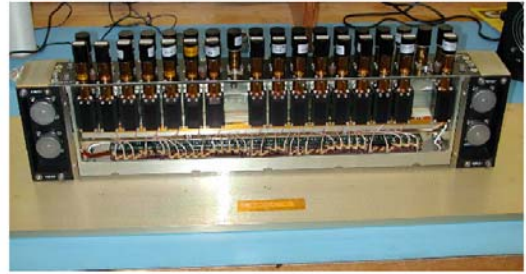

Figure 11. ACD electronics chassis: Front End Electronics (FREE) board is under the photomultiplier tubes (PMTs – black cylinders). Tops of the PMTs are covered by black caps, to be removed during mating of the fiber cables to the PMTs. The High Voltage Bias Supplies (HVBS) are located under the FREE board

Four FREE boards are mounted on each of two opposite sides of the ACD, containing a total of 144 channels; the remaining two sides contain two FREE boards each, with a total of 72 channels; only 194 of the 216 channels are used. One (of two redundant) High Voltage Bias Supply (HVBS) powers the PMTs associated with each FREE board. Each FREE board connects directly to the GASU, with no interconnection between ACD FREE boards.

Design of the ACD electronics was strongly affected by the strict mechanical requirement to fit the FREE boards and PMTs within very limited space. This requirement strongly affected the PMT choice (its diameter and length) and PMT positioning relative to its FREE board. One of the consequences of this is very small gaps between adjacent PMTs, which created significant challenges in routing the fibers and assuring good optical coupling of the fiber cables to the PMTs (Fig. 11).

### 3.11.1. Front-end electronics

The Front End Electronics (FREE) has interfaces to the GASU, the High Voltage Bias Supply (HVBS), and the PMTs. The FREE circuit board includes 18 analog Application Specific Integrated Circuits (ASICs), 18 analog-to-digital converters (ADCs), and one digital ASIC. The analog ASIC, designed at SLAC, provides amplification, discrimination, shaping, and sampling of the signal from the PMT. It was fabricated in the Agilent 0.5um CMOS process and packaged in a 44 pin 10.0 mm x 10.0 mm TQFP plastic package with leads on a 0.8 mm pitch. There are three discriminators for each analog channel. The VETO discriminator is used for signaling passage of any charged particle. The high-level discriminator (HLD) is used for signaling heavy nuclei passage (high amplitude signals). The third discriminator is used to control the choice between the two gain ranges of the pulse height digitization. Although the principal purpose of the ACD is to identify charged particle passage (VETO), a secondary mode provides a signal (HLD) of a heavy particle, used as a calibration mode for the LAT calorimeter. In order to span the range from 0.1 mip to hundreds of mip, the pulse height analyzer operated in two gain ranges.

The digital ASIC, designed at Goddard, provides the interface to the eighteen analog channels and contains the digital logic and interface circuitry required to prepare event data for transfer to the GASU. It has the following functional blocks: GASU interface, 18 analog ASIC interfaces, command processor, data formatter, PHA control logic, VETO delay logic, and hitmap logic. The GASU interface consists of differential drivers and receivers that are based on the standard LVDS protocol, a 3.5 mA current-to-voltage driver-receiver set. The analog ASIC interfaces utilize a custom-designed current-mode interface used to communicate to the 18 analog ASICs. The command processor receives command information from the GASU and processes valid commands to change the state of the FREE circuit card. The data formatter is a state machine operating on a command-response protocol that formats and transmits serial data to the GASU upon a command request. The PHA logic is responsible for controlling the 18 ADCs on the FREE board, capturing the digital pulse height information, and providing digital zero-suppression functions, if requested. The trigger delay logic provides for a commandable delay between the Trigger Acknowledge (TACK, from the LAT trigger electronics) and the shaping amplifier hold signal sent to the analog ASICs. The hit map logic provides all combinatorial and shift-register delay functions associated with providing a serial hit map of which PMT signals exceed the VETO threshold within a commandable time window. The hitmap, generated from relatively fast discriminator signals, is much less susceptible to pile-up than the much slower PHA process, and is used to compensate for such pile-up (see section 4.3 for details). The digital ASIC was fabricated in the Agilent 0.5um CMOS process and packaged in a plastic quad flatpack.

The ACD utilizes a commercial analog-to-digital converter from Maxim semiconductor, a MAX145 12-bin serial ADC in Maxim's 8-pin μMAX plastic package, suitably qualified and screened.

### 3.11.2. High voltage bias supply

The low power HVBS uses a simple, reliable and flight-proven circuit approach. The voltage is generated from a 700 kHz resonant sine wave Hartley oscillator with a 170 volt peak-to-peak output. The ten-stage multiplier steps the voltage to a maximum of 1300 volts. The output voltage is adjusted by a 0-2.5 V control. The HVBS incorporates a planar transformer, and output regulation is achieved by controlling the conduction of the oscillator switching transistor. Each HVBS provides power to 15-17 PMTs, all associated with the same FREE circuit card. In each PMT divider assembly, a large series resistance prevents a short-circuit in that PMT divider from dragging down the voltage available to the other PMTs.

### 3.12. Tile gap analysis and the ribbons

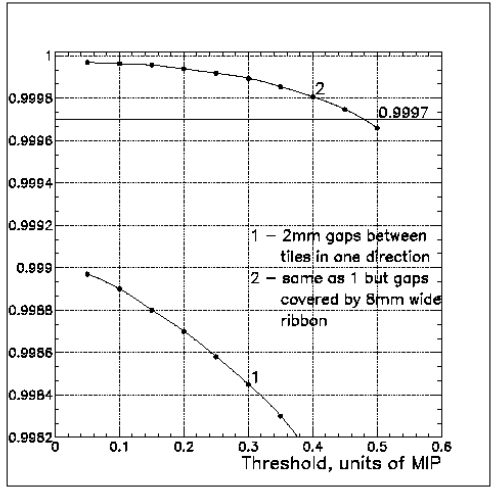

Figure 12. Simulated ACD expected performance and effect of the ribbons. Line 1 – ACD efficiency without ribbons, and line 2 – with the ribbons. Hereafter on similar efficiency plots the lines are shown to guide the eye

The segmented ACD provides the capability to suppress the backsplash effect and extend LAT useful detection of gamma rays up to several hundred GeV. It also provides tolerance to micrometeoroid penetration through the wrapping of a tile. However, it also necessitates unavoidable gaps between tiles. Design of the ACD assumes the presence of 2-3 mm gaps between the tiles for the wrapping, tile thermal expansion and prevention of tiles hitting each other when vibrations occur. The ACD layout overlaps tiles by 20 mm in one direction, providing essentially 100% coverage in that dimension (a trade study convinced us that a design with overlaps in two dimensions would add too much complexity, mass, and volume).

The effect of these gaps on the ACD efficiency has been studied in depth. It was found in the simulations that 2 mm gaps between tiles in one direction (tiles are assumed overlapping in other direction) immediately pushes the efficiency of the ACD below the required level of 0.9997 (Fig. 12, line 1).

The solution was to cover the gaps by flexible scintillating ribbons. The gaps at the ACD corners are not protected by ribbons; however, the small fraction of events entering through these corner gaps and triggering LAT will necessarily be along the LAT periphery, and they can be removed during analysis. The scheme of tile layout with the ribbons is shown in Fig. 13.

In order to provide flexibility in the ribbons, they were made of 1.5 mm square scintillating fibers with ~60 μm thick double cladding, provided by Washington University, St. Louis. Each ribbon contains 25 separate fibers in three layers, approximately 3 meters long (Fig. 14). First the fibers of each layer are bonded together with Uralane, using 50 μm thick aluminum foil as a substrate. After that, the three fiber layers are bonded together with a 1/3-fiber offset, needed to eliminate "channeling" of particles along the cladding, clearly seen in efficiency simulations. The next step is to shape the ribbon to conform to the profile of tile staggering. This is very delicate step, because it is very easy to degrade the scintillating and light-conducting properties of the fibers during shaping. A special bending fixture was made, with the exact ribbon profile to be shaped, and the entire setup was placed in an oven for 10 hours to shape the ribbon. The oven temperature was kept in the range 50-55 C so as not to damage the cladding. Eleven ribbons were fabricated, eight for the flight unit and three spares. Every ribbon has a slightly different shape depending upon its position in the ACD.

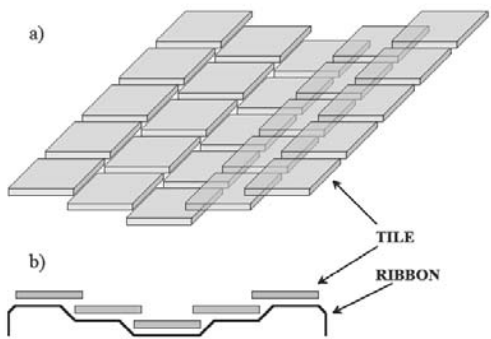

Figure 13. 3D schematic of tile overlap (a) and cross section (b) for the top of ACD

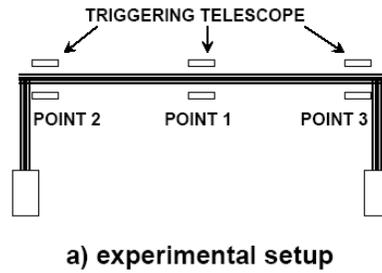

a) experimental setup

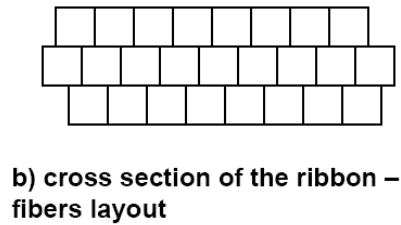

b) cross section of the ribbon – fibers layout

Ribbon light yield from cosmic muons was measured at three points (see test setup in Fig. 14): at the ribbon center (point 1), and two points near the large bends (points 2 and 3). The triggering muon telescope, consisting of two 1 cm square plastic scintillators was used to trigger pulse height analysis of signal due to muons crossing the ribbon at the given point. The number of photoelectrons was estimated by comparing with the signal from a reference tile, obtained with the same PMT with gain of $1.1\times10^7$ and photocathode quantum efficiency of 0.145.

Figure 14. Ribbon test setup (a) and ribbon cross section (b)

It was found that the signal from the ribbon center (point 1) is ~4.5 photoelectrons, and a signal of approximately 7 photoelectrons was obtained at points 2 and 3 for the corresponding nearest PMT. This agrees with the signal expected from *mip* light production in the ribbon (4.5 mm of plastic scintillator), light attenuation measured in a similar test using of a 3.5 MeV (endpoint) β–source $^{106}$Ru. The measurement of attenuation length is shown in Fig. 15, line 1, demonstrating a gradual increase of attenuation length from ~120 cm within 1 m from the light source, to ~160 cm further away.

Fig. 12, line 2, presents the result of preliminary simulations showing how the ribbons solve the gap problem. Ribbon light yield was taken to be 4 photoelectrons, and the detection threshold was assumed to be 2 photoelectrons. Detailed Monte Carlo simulations along with mechanical considerations resulted in the final design values for ACD detectors gaps. Gaps at room temperature between adjacent tiles and vertical clearance are 2 mm, while gaps between tiles of different planes at the corners are larger, from 3 to 4 mm. One of the principal goals of the ACD integration was to maintain these gaps per design. Final mechanical checks demonstrated in most cases ±0.5 mm agreement between real and designed gaps.

3.13. Light attenuation in the WLS fibers and the resulting need for clear fiber cables

Since the ACD tiles are positioned at different distances from the PMTs, with the largest distances exceeding 1 meter, the need to maximize the light transmission to the PMTs means light attenuation becomes an issue. Two options were considered: 1) deliver the light from tile to the PMT via the same WLS fibers that are used to collect the scintillation light in the tile; and 2) couple WLS fibers to clear fibers with longer attenuation length. Light attenuation properties were measured for WLS fiber BCF91A and for clear multiclad fiber BCF98, which can be used in clear fiber extension cables (Fig. 15, lines 2 and 3). The attenuation length in clear fiber was found to be ~6 m, (somewhat shorter than the manufacturer's specification). The attenuation length in the WLS fiber was found, as for the scintillating ribbon fibers, to increase from ~120 cm within the first 50 cm from the light source to ~4 m at a distance of 1 m.

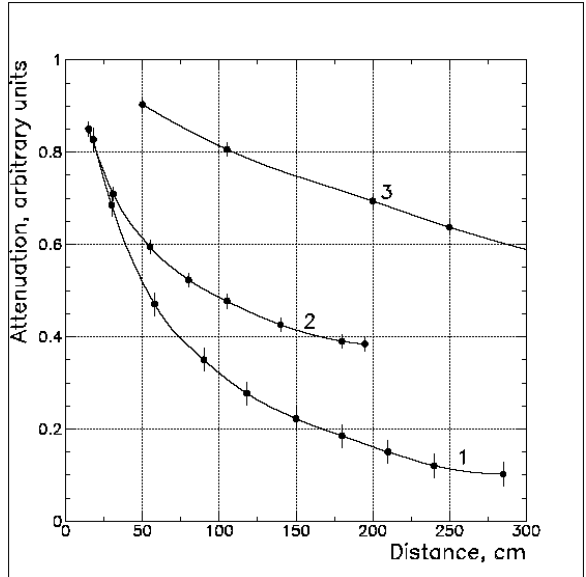

Figure 15. Attenuation length for 1) ribbon fibers, 2) WLS fibers, and 3) clear fibers

A highly efficient optical connector was designed to couple the 1.0 mm diameter WLS fibers to 1.2 mm clear fibers. The WLS end of a connector can be seen in Fig. 6. The larger diameter of clear fibers provides tolerance for fiber misalignment in the connector. The light transmission efficiency of this connector was found to be 85%. It was found that for fiber runs of less than 50 cm from the tile to the PMT, the optical connector is not needed; for longer runs, the optical connector provides better light delivery to the PMT. ACD mechanical constraints put some restrictions on fiber cable routing, especially for the 3$^{rd}$ row side tiles, where the faces of the PMTs are only 4 cm below the tile edges. The final result is that all tiles on the ACD top and the two upper side rows have clear fiber cable extensions and all 3$^{rd}$ and 4$^{th}$ row tiles have WLS fibers running all the way to the PMT locations. As a result, the total light loss from exiting a tile to arriving at the PMT varies from 20% to 60% depending on fiber cable length.

3.14. Photomultiplier tubes (PMTs)

The top level requirements for the ACD PMTs are:

- robust to survive launch and space flight conditions,

- gain and noise characteristics suitable for detecting small numbers of photoelectrons,

- physically small, because the ACD requires 194 PMTs in a very limited volume.

The tube selected is the Hamamatsu R4443, a ruggedized version of the R647 that was used for the 1997 beam test. The R647 tube (the non-ruggedized version) was used on the RXTE satellite [14], and a nearly identical R4444 was used successfully on the GOLF instrument aboard the SOHO satellite [15]. Hamamatsu kindly optimized the quantum efficiency (QE) at 490 nm, the peak of the WLS fiber transmission spectrum, rather than the normal 420 nm. This improved the average QE by 10-15%, so the QE for the 194 flight PMTs ranges from 16% to 23%. The specified gain of the PMTs was $2 \times 10^6$ at the maximum voltage of 1250 V. All the tubes delivered by Hamamatsu exceeded this requirement.

PMT gain is known to change over time; the rate of change depends mainly on the anode current. Although all tubes of a given type eventually decrease in gain at roughly the same rate, the initial change varies considerably; indeed, some R4443 PMTs increase in gain initially, although in those the gain change eventually reverses. This pattern of change must be taken into account in the ACD design, because a single HVBS feeds 15-17 PMTs, and there is no capability in the electronics to compensate for individual PMT gain changes. The high voltage can be raised to correct for a change in *average* gain, but there is no way to accommodate *divergence* of the signals from such a group of PMTs. To provide a means to compensate for this variation, Hamamatsu provided initial gain change test data for all flight PMTs. All tested PMTs demonstrated their gain to be within the range from 0.6 to 1.0 of that before the test, after the 100-hours accelerated burn-in test, corresponding to approximately 4 years of normal operation in an orbit. The ACD initial PMT HV settings (700-800V) leave us with ample margins to tolerate long-term PMT gain degradation

Because of unit-to-unit variations in PMT gain (up to factors of 20), in quantum efficiency (up to 40%), and in variations in gain change mentioned above, as well as variations in light production and transmission to the PMTs (different tiles and different length clear fiber cables), assignment of PMTs was a tedious process, roughly as follows:

1. The light signal from each fiber cable was estimated, using the tile thickness and the fiber cable light transmission characteristics. In general, the long tiles have lower light signals than the smaller tiles, and the ribbons have even smaller signals.

2. PMTs were sorted by gain change rate, so that tubes with similar rate could be assigned to a particular HVBS.

3. Within such a group, the highest *gain* PMTs were assigned to the ribbons, so that their PMT signals would be very roughly similar in magnitude to those from tile PMTs. For the same reason, PMTs for the long tiles were chosen to have rather high gains.

4. For the smaller tiles, the fiber cables providing smaller light signals were assigned to PMTs with high QE; those with larger light signals use PMTs with lower QE.

5. Because of limited spare PMT availability, there were inevitably cases where two or more fiber cables would be optimized with the same PMT. In these cases, "important" tiles (those in the ACD top and the first side row) were given preference. For the third row tiles, the PMTs available were sometimes substantially non-optimal.

Mechanical design of the PMT housing was a challenging task as well. The requirements are as follows:

a) Very little room is available for PMTs near the FREE boards: 30 mm PMT center-to-center (see Fig. 11);

b) The housing must be light tight;

c) PMTs must be protected from mechanical stress from the housing due to vibration and shock, and also over the survival temperature range −40 C to +45 C;

d) Over the same temperature range, the housing design must provide reliable optical coupling of fibers to the PMT window with maximal light conductivity; it must be easily disconnected and reconnected if necessary;

e) The housing must shield the PMT from a magnetic field of 2 G, to less than ~0.2 G inside.

f) Voids in the PMT housing and voltage divider chain enclosure must be adequately vented for launch without compromising light-tightness;

g) The mass of the 194 housings must be minimized.

A design in which the PMT was supported in silicone potting proved to be untenable because of the close spacings and very diverse coefficients of thermal expansion. The final design utilizes a mounting in which the PMT is captured between semisoft plastic parts inside a machined aluminum housing (without potting). Because of irregularities in the PMT shapes, the plastic parts were required to be hand-fitted to each of the 194 tubes (plus spares). The tube housing is threaded to allow light-tight attachment of the fibers, which are glued into a bushing. The housing is wrapped in mu-metal for magnetic shielding. The resistive divider chain is implemented on a twice-folded multilayer flexible printed circuit, mounted in a vented light-tight box 27 mm × 27 mm × 46 mm. Flight-approved black tape is used on seams to provide additional protection against light leaks.

### 3.15. Simulations of ACD expected performance

After design completion and before moving to the fabrication phase, we had to validate the ACD ability to detect charged particles efficiently. The approach was to simulate the ACD mechanical design as closely as possible, then to insert measured or derived detector performance parameters, and then calculate the resulting ACD efficiency for singly charged particles. This allowed us also to determine margins in case of failures during a long mission.

In the ACD simulation we took into account the fact that all mechanical gaps will open up in orbit due to temperatures well below room temperature causing thermal contraction of tiles and other materials. Fig. 3 shows the model used in the Geant-3.21 simulations. The ACD geometry was simplified slightly by replacing gentle fiber ribbon bends with 90° corners, and also the 2nd and 3rd rows of side tiles are not tilted in the simulation. The former approximation makes the simulation results more conservative, because it creates larger openings between ribbon and tile. The latter approximation is not expected to affect the estimated ACD performance because it does not change the gaps between detectors. The results of the simulations are shown in Fig. 12, line 2. These results were preliminary, because the most important parameters affecting the performance, tile and fiber ribbon light yield, and mechanical gaps, were taken as per-design. Conservative values of 20 photoelectrons were used for the tile light yield, and 4 photoelectrons from the ribbon center, assuming that the true values (to be obtained after ACD integration) would be better. The simulations demonstrate that ACD "per-design" meets the requirement to achieve 0.9997 average efficiency, with the Veto threshold set to 0.3 *mip,* over the entire area. The simulations also included the case of a single FREE board (or HVBS) failing (which would mean 15-17 ACD detecting elements, tiles or ribbons, running with only one PMT), and the case of general light yield reduction by 25%. It was found that ACD performance has adequate margins for those failure cases. Special simulation runs were performed to understand the effect of the mounting holes, which are effectively the dead spots in the tiles. It was found to be insignificant: 4 holes in each ACD tile reduce the whole ACD efficiency by ~ $3\times10^{-5}$ for the isotropic flux. Finally, to understand the available margins on the gaps between tiles, simulations were done varying those gaps. The most critical gaps were found (as expected) to be at the ACD corners (~ 4mm), where the gaps are not covered by ribbons. Their increase by only 20-25% would significantly degrade the ACD performance. The back-up option is to eliminate that area in off-line analysis.

Even after completion of the ACD, a direct measurement of the high efficiency required over the large area of the ACD without the tracker and calorimeter to provide direction and energy information was impractical, particularly under the temperature range expected in space. For this reason, the acceptance test of the ACD also used the same type of simulation process (see section 5.2.2).

4. ACD performance and operational issues

This section discusses specific aspects of ACD operation that have been investigated during the ACD design.

4.1. Broken fibers

Wavelength shifting fibers are a key element in ACD, providing light collection from the scintillator tiles and delivering the light to the PMT. The fibers are made of polystyrene, with a thin polymethylmethacrylate (PMMA) cladding layer for improving light containment inside the fiber. Undergoing numerous bends both inside and outside the tiles, they have potential for kinking or even breaking. To reduce light conduction in a fiber dramatically, it is not necessary to break it completely; improper fiber bending can easily break the fiber cladding and dramatically increase light escape from the fiber.

In order to understand ACD tolerance to broken fibers, special tests were performed. *Mip* peak positions (proportional to the tile light yield) were measured for a tile with all fibers in place, and after that fibers were removed gradually to see the light yield degradation. It was concluded that up to four broken fibers per tile can be tolerated, as long as the broken fibers are separated by at least two good fibers. The expected light yield degradation, averaged over the tile area, does not exceed 10%, but there is somewhat larger light yield reduction in the vicinity of broken fibers.

Although some fibers were damaged on one of the third row tiles during assembly, there is no indication of any broken fibers on any of the other tiles.

### 4.2. PMT operation at high event rate

The design of the PMT resistive divider minimizes the power drawn from the HV power line. The conceptual design of the divider is standard: a series of resistors to distribute the HV between PMT dynodes, with the last three dynodes supplemented by capacitors to provide extra charge in the last stages. This current is minimized to the value necessary to assure the PMT operation, without signal degradation, at the expected cosmic ray signal rate.

The maximum expected rate of singly charged particles through a single ACD tile (excluding the long tiles) is 1.-1.5 kHz; the system is required to handle a 3 kHz rate without performance degradation. The divider is designed to provide ~1 µA at 800 V, a factor of 300 larger than the maximum expected PMT output current in orbit. The PMT was tested with a pulsed LED placed at the tube face, producing four times larger signals than expected from a *mip*. The PMT was observed to saturate at a rate of ~100 kHz. This is consistent with expectations and provides confidence in PMT operation at high rates with large margins.

### 4.3. Electronics signal pile-up

Signals from the low-power analog ASICs used for processing the PMT outputs must be analyzed with some care, because they do not handle signal pile-up cleanly. The VETO discriminator signals used in the on-board processing have a fixed timing width and a relatively high threshold, in order to be used in the hardware trigger. This signal width is smaller than the time needed for the comparator circuitry to recover, so that a second signal coming shortly after the first will not produce a second VETO pulse. This pile-up limits the ACD efficiency to a level acceptable for the on-board processing but not for the full 0.9997 requirement.

The ground processing uses the "OR" of two other signals in order to achieve the full efficiency:

1. The PHA signal from each phototube is sent with each LAT trigger. In principle, the threshold can be optimized to achieve the required efficiency. This PHA, however, has a shaping time of ~4 µs, required for compatibility with the LAT data acquisition circuitry, and this signal has a significant overshoot before recovery. This overshoot creates the possibility of pile-up inefficiency for signals separated in time by less than 10 μs (especially if the first one is caused by a particle with Z>1). The effect of pile-up on efficiency was simulated for an average particle rate of 1 kHz and natural cosmic ray composition of protons and helium. The results are shown in Fig. 16, line 2, compared to line 1, where pile-up was not considered. It is seen that pile-up immediately causes the ACD efficiency to fall well below the required level.

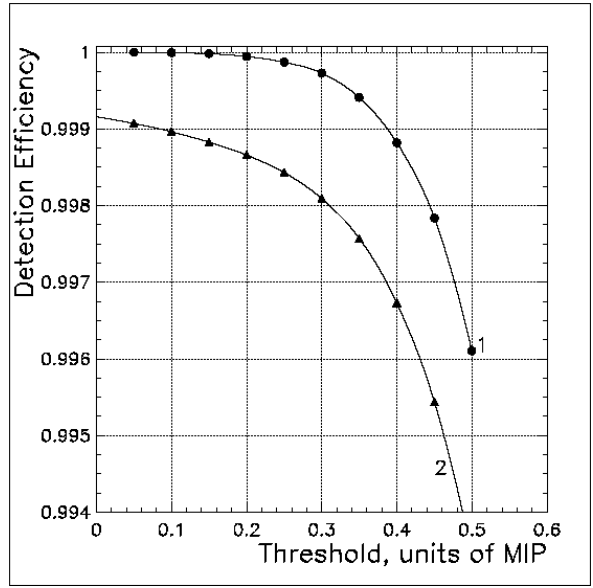

Figure 16. ACD simulated performance with pile-up effect: 1) not included, and 2) – included

2. A second, length-extended Veto signal, covering the signal time-over-threshold and baseline recovery time, solves this problem. This "hitmap veto" is basically the same signal that is supplied to the LAT trigger electronics, but with variable length to cover any event arriving within the PHA pile-up interval of the previous event or the dead time after the fixed-width veto signal. It has a threshold that may be too high to achieve (alone) the required efficiency, but the probability of having two events close together in time (which would produce PHA inefficiency) AND having a signal fluctuation below the discriminator threshold (which would produce discriminator inefficiency) is extremely small. Simulations were done with various Veto thresholds (0.4, 0.5, 0.6 *mip*) and an event rate of 1.15 kHz; this demonstrated that the efficiency was restored to line 1 of Fig. 16. Increasing the signal rate to 4 kHz did not cause significant changes.

### 4.4. Thermal cycling

Because of the broad operational temperature range expected for the TDAs, thermal issues are very important during ACD operation on orbit. The TDAs are passive from the point of view of heat generation, and most are located at some distance from heat-emitting electronics. Although they are covered by the MMS and thermal blanket, they will be the portion of LAT most affected by temperature variation in orbit. A test was performed to determine whether there is any light yield degradation due to the thermal cycling. Such degradation could potentially be caused by cracking of the optical glue embedding the WLS fibers in the tile grooves, by degradation of the optical coupling between a fiber cable and its PMT, by separation of the fiber cladding from the fiber core, etc. Two identical tiles were fabricated; the first one was kept as a reference, and the second experienced thermal cycling, 440 cycles over the range −60 C to +45 C. No significant light yield change was found within the measurement accuracy.

### 4.5. Thermal variation of ACD signals

Two properties of the ACD detectors are known to have substantial temperature coefficients: the larger contribution is the PMT gain, and a smaller contribution from the scintillator light output. The temperature dependence of the WLS fiber conversion efficiency and light transmission of WLS and clear fibers were not known. In order to characterize the overall temperature TDA performance, three tests were done over −60 C to +45 C, in 5 C steps, measuring the relative amplitude of the *mip* signal. The results are summarized in Table 1 for sample units.

Table 1. Thermal variation of the amplitude of *mip* signals from ACD components

| Test | Components outside temperature chamber | Components inside temperature chamber | Average temperature coefficient, %/°C |
|---|---|---|---|
| 1 | PMT, optical connector, clear fiber cable, part of WLS fiber cable, | Tile, part of WLS fiber cable | −0.1 |
| 2 | PMT | Tile, WLS fiber cable, clear fiber cable, optical connector | −0.35 |
| 3 | None | Tile and PMT, no optical connector | −0.5 |

Hamamatsu reports that the thermal coefficient for a typical PMT is ~ −0.2 %/°C, a value consistent with our measurements. Note that all of the measured temperature coefficients are negative, so signal increases at lower temperatures. Note also that signal changes due to the tile and the optical connector cause changes in light yield; while for the PMT gain changes affect only signal amplitude. The PMT QE is known to be temperature stable, so the PMT changes do not affect the detection efficiency, as along as the discriminator threshold is adjusted appropriately. From the results obtained we expect the temperature change of the signal from the combination (Tile + optical connector + PMT) to be ~ −0.75%/°C. We performed more detailed thermal tests with the assembled ACD, confirming this result (section 5.2.2), and we estimate that the ACD Veto threshold settings have to be readjusted whenever there is a 10 C temperature change. Taking into account that the expected ACD operational temperature in orbit will be 0 C and below, we will gain significant light yield over that measured in the laboratory.

5. Integration, tests and calibration of the ACD

### 5.1. ACD integration

The ACD integration was a delicate process, requiring careful sequencing of operations. All of the integration steps were tested on prototypes before moving to the flight structure. The final appearance of the ACD (before it was covered by the MMS) is shown in Fig. 17.

1. The ACD mechanical structure was assembled and equipped with tile and ribbon mounting flexures, FREE chassis attachment brackets, 44 thermistors and 9 accelerometers.

2. The ribbons were installed and fastened to stand-offs in a manner that allowed for thermal movement.

3. The tiles on the ACD top were installed (starting with the middle row), maintaining ~2 mm gaps between adjacent tiles, and vertical clearance between overlapped tiles of 0.5-1 mm from tile wrapping to wrapping. Ribbons were centered under the gaps between tiles, and vertical clearance between ribbons and tiles were set to 0.2-0.5 mm to allow free ribbon thermal movement.

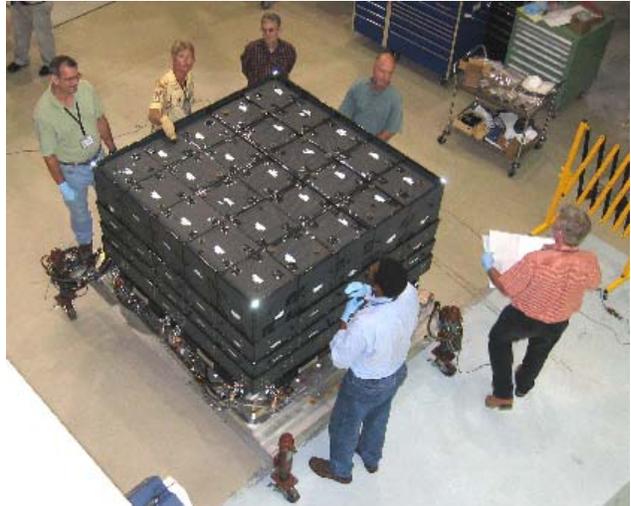

Figure 17. The ACD assembled!

4. Side tiles were installed by rows, starting with the top row. All 20 tiles of the row (5 tiles on each side) were installed, maintaining the specified tile to tile gaps. Special care was given to the corner gaps, where top tiles meet with side tiles, and where two side tiles meet at a corner.

5. After each row of tiles was installed, its fiber cables were routed and connected to the assigned PMTs, and tested for light tightness. This assembly procedure was repeated for each of the three side rows.

6. Light tightness testing was performed on each side after its installation was complete.

7. The final detector integration step was the installation of the bottom (long) tiles (which cover the PMTs and electronics), including connecting them to their PMTs and testing for light tightness.

8. After final testing was performed, the ACD was covered by the thermal blanket and micrometeoroid shield.

## 5.2. ACD ground tests

### 5.2.1 Functional and performance tests

Taking into account the extreme importance of instrument reliability and sustainability for a long space mission, the task is not only to test the ACD operation. In addition, it is necessary to understand operational margins, and consequences and mitigation of possible marginal operation if it occurs.

The functional test package includes the following:

- voltages, currents, temperatures
- commanding
- electronics performance
- detector operation
- timing parameters
- margin testing (for voltages and clock frequencies)

Test Charge Injection (a capability of the analog ASIC) was used to test and calibrate the performance and linearity of the flight electronics, after which the flight electronics were used to test and calibrate the TDAs, using the VETO and hitmap discriminator readouts, along with the PHA data, all of which are available in the data stream. . The first parameters to be determined were the ACD PHA pedestals, followed by the main task of the ACD performance test, determining the *mip* peak position in the pulse-height histogram for every ACD channel (PMT). Knowledge of *mip* peak positions allows us to determine the proper Veto threshold settings (in units of *mip*).

*Mip* peak determination utilizes normally (or nearly so) incident particles because otherwise it is affected by off-normal events whose longer path lengths in the scintillator produce larger energy depositions in the tile. During ACD standalone tests we use an ACD self-trigger, created by coincidence of signals between any two ACD sides, or one side and the top. In the data analysis we select events for every tile by requiring that they are coincident with a pre-determined set of tiles to select quasi-normal incidence particle events.

Tile light yield was determined by using external scintillating hodoscopes above and below the tile under test, and the analysis approach was similar to that described in section 3.4 and Fig. 8 and 9. This was the first time the tiles were operating with their assigned clear fiber cables and PMTs, and with flight-configured routing of all cables, which affects the light propagation through them. The light yield values obtained exceed those expected by an average of ~10%. Ribbon performance was tested by using the external hodoscope to determine the *mip* peak from both PMTs at several places along the ribbon.

### 5.2.2 ACD performance verification

The fully integrated ACD experienced the normal battery of tests for space flight hardware: vibration, acoustics, and thermal vacuum. Stability of all parameters was confirmed to be within the required range. Thermal dependence of the *mip* peak positions was found to be ~0.8%/°C in the temperature range from −25 C to +35 C, which is consistent with that given in section 4.5. Pedestals are much less affected by temperature. These measurements provided information about the required precision of temperature monitoring. As was expected, thermal behavior of *mip* peak positions imposes the most stringent requirement on their calibration: in order to provide veto threshold settings accurate to ~0.05 *mip*, positions of *mip* peaks must be known for every 10 C temperature change.

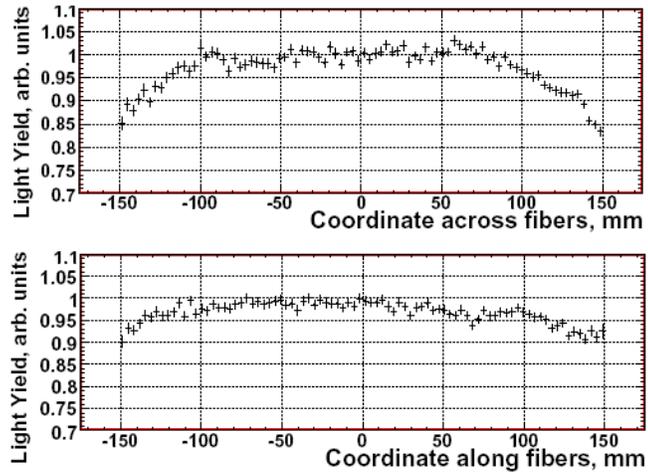

Figure 18. Tile light collection uniformity (in relative units). Upper panel shows light collection uniformity across the tile WLS fibers, and lower panel – along the fibers

After integration of the ACD into the LAT, event selection capabilities of the tracker and calorimeter were used to calibrate the ACD accurately. Using precise track reconstruction from the data collected in ground muon LAT tests, normally incident muons were selected for every tile and ribbon. Signal yield was determined (in photoelectrons) for each tile/PMT combination. The resulting numbers exceed those expected and have an average value of 23.5 photoelectrons. LAT tracking was also used to study light collection uniformity over tile area (Fig. 18). The results meet the requirement of <10% light collection non-uniformity over each tile area, and <30% light yield degradation within 3 cm of a tile edge. Light yield along fiber ribbons was also measured with precise tracking. Attenuation in the ribbon was found to agree with that measured in the laboratory, shown in Fig. 14, line 1. The light yield in the ribbon centers was found to be 3.8–4.7 photoelectrons, which also agrees with expectations. This confirms that using ribbon thresholds of 1.0-1.5 photoelectrons provides adequate detection efficiency.

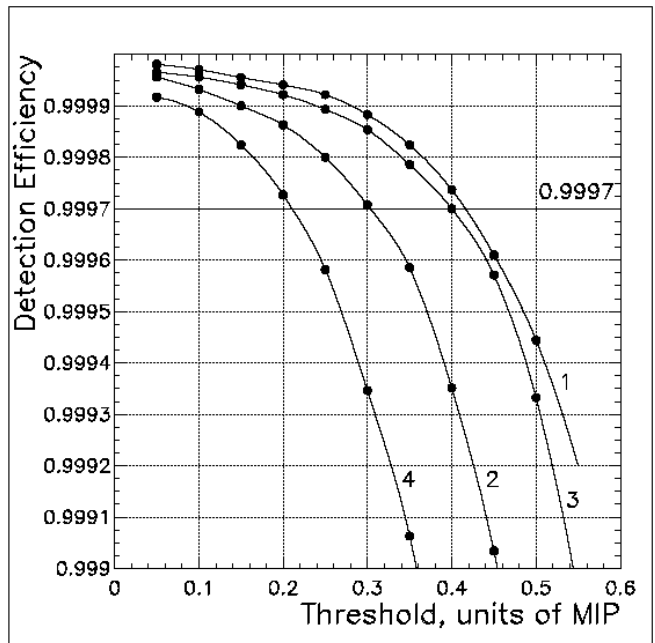

Figure 19. Final ACD detection efficiency. 1 – Nominal operating condition; 2 – 1 FREE failed; 3 – 60% of the nominal light yield; 4 – 1 FREE failed and 60% of the nominal light yield

All ACD performance parameters, such as *mip* peak positions and light collection uniformity for tiles, and light yield along the fiber ribbons, were placed in the simulation model to calculate the resulting overall ACD efficiency for detection of singly charged relativistic particles. The same model as described in 3.15 was used for this purpose, but updated with the newly-obtained ACD parameters. Improved light yield and light collection uniformity values compared with those used in the design phase resulted in higher ACD efficiency, providing significant margin. In order to account for the light yield determination uncertainty, 20% light yield fluctuation was applied to each value. If operated with 0.3*mip* Veto threshold, with light yield (from all tiles and ribbons) reduced to 60% of the original, or with one FREE board failed (up to 17 tiles and ribbons operating with a single PMT instead of two), the ACD still meets the efficiency requirement (Fig. 19). With a reduced Veto threshold (0.2 *mip*), the ACD could meet its requirements with both failure conditions simultaneously (loss of one FREE card and 60% of the original light collection).

Verification of the backsplash requirement was based on predictions from the beam test. The expected LAT efficiency degradation caused by backsplash is shown in Fig. 10 (for the events entering the instrument through the top), demonstrating that the ACD exceeds the backsplash requirement. Having this margin on the backsplash allows the possibility of operating with a lower veto threshold to meet the efficiency requirement if necessary.

6. Conclusions and acknowledgements

As a result of approximately eight years of effort, the LAT Anti-Coincidence Detector has been successfully designed, built, tested, and integrated into the LAT. The ACD array of plastic scintillator tiles with wavelength shifting fiber light collection plus scintillating fiber ribbons and photomultiplier tube readout meets all the science performance and spacecraft reliability requirements for the GLAST LAT. Currently LAT is in environmental testing, after which it will be integrated into the spacecraft and finally delivered to orbit.

During the years of ACD activity more than 100 people have been involved in the design, fabrication and test processes. We are grateful to Steve Ritz, LAT Project Scientist, for his constant interest, support, and valuable suggestions throughout the ACD development. Early in the ACD program, we benefited greatly from consultations with Don Stillwell, Andy Dantzler, Jay Norris, and John Mitchell of GSFC, and also Pavel DeBarbaro of Fermilab. Bill Atwood of UCSC provided important design and performance testing suggestions. The key ACD detector, the TDA, was fabricated at FermiLab by the team of Phyllis Deering (the lead), Todd Nebel, John Korienek, John Blomquist, Eileen Hahn and others. Crucial contributions were made to ACD design and fabrication by Art Ruitberg and the HVBS team, Bob Baker and Norm Dobson,

Glenn Unger, Satpal Singh, Chuck Hanchak and the electronics team, Russ Rowles and the composites team, Bob Binns and his team at Washington University (fibers for ribbons), Carlton Peters and the thermal team, Fred Gross, Pilar Joy and the materials team, Ron Kolecki, Tavi Alvarez, Al Lacks and the quality assurance team, George Shiblie, Steve Schmidt, Ian Walker, Jim Woods, Brian Grammer and the mechanical team, Thom Perry, Nick Virmani, Lou Fetter and the parts team, Diane Schuster and the MMS team, and graduate students David Wren, Luis Reyes, and Chris Stark. Jim Odom deserves special mention for his invaluable work on both electronics and integration and test.

We appreciate very much the crucial software effort by Karen Calvert, Sharon Orsborne and Cameron Jerry, and everyday software help from Heather Kelly and web master J.D. Myers. Under often frustrating conditions, management support was cheerfully provided by Deanna Adamczyk, Lorna Londot, Andy Eaker, and Maxine Windhausen. Special thanks go to the heroic efforts of the Integration team led by Craig Coltharp, Allen Crane and Darian Robbins, and the test conductor team led by Adrienne Beamer and George Moore.

Completion of the ACD would have been impossible without dedicated and high quality work of our technicians Deneen Ferro, Bill Daniels, Traci Pluchak, Bahe Rock, Paul Haney, Steve Harper, Nick Kwiatkovski, Dawn Blackburn and others.

A number of important contributions to the ACD were provided by Stanford Linear Accelerator Center personnel. Gunther Haller, Michael Huffer, Oren Milgrome, and the electronics team were vital in defining and building electronics interfaces. Richard Claus, James Panetta, and the online software team offered vital support for the test team. Critical contributions in the last phases of ACD testing and instrument performance analysis were provided at SLAC by Eric Charles, Eduardo deCouto e Silva, and Anders Borgland.

And finally we want to express our gratitude to Eric Charles, Robert Johnson and other LAT team members for careful reading the paper and providing many of valuable comments.